\documentstyle[psfig,prb,aps]{revtex}
\begin{document}

\twocolumn[\hsize\textwidth\columnwidth\hsize\csname
@twocolumnfalse\endcsname

\title{
{\em Ab initio} calculation of the lattice distortions induced by 
substitutional Ag$^-$ and Cu$^-$ impurities in alkali halide crystals.
}
\author{Andr\'es Aguado, Jos\'e M. L\'opez, and Julio A. Alonso}
\address{Departamento de F\'\i sica Te\'orica, Facultad de Ciencias,
Universidad de Valladolid, 47011 Valladolid, Spain}
\date{}
\maketitle
\begin{abstract}
An {\em ab initio} study of the doping of alkali halide crystals
(AX: A = Li, Na, K, Rb; X = F, Cl, Br, I) by ns$^2$ anions (Ag$^-$ and
Cu$^-$) is presented. Large active clusters with 179 ions embedded in
the surrounding crystalline lattice are considered in 
order to describe properly the lattice
relaxation induced by the introduction of substitutional impurities. In all the
cases considered, the lattice distortions imply the concerted movement of
several shells of neighbors. The shell displacements are smaller for the
smaller anion Cu$^-$, as expected. The study of the family of rock-salt
alkali halides (excepting CsF) allows us to extract 
trends that might be useful at a predictive level in the study of
other impurity systems. Those trends are presented and discussed in terms of
simple geometric arguments.
\end{abstract}

\vskip2pc]

\section{Introduction}
\label{aguado:intro}

Most of the luminescent materials presently used in several technological
applications\cite{Bla95} involve the doping of a pure ionic crystal, that is 
substitution of some of the ions by other ions with specific
absorption-emission characteristics. 
The fine details of the absorption-emission spectra, as well as the efficiency
and resolution of the scintillator, are determined by the system-specific
embedding potential acting on the impurity, which is in turn sensitive to the
distortion induced by the impurity on the crystal lattice. Thus, a theoretical 
understanding and accurate determination of those distortions is of
paramount importance, moreover if we realize that their experimental measurement
is a difficult task.\cite{Bar92,Pon90,Zal91}

Two main methods are applied nowadays to model impurity systems: supercell
techniques, that exploit the convenience of the Bloch theorem by
periodically duplicating a finite region of the crystal around the impurity
\cite{Pus98}; and the cluster approach, in which the doped crystal is modeled
by a finite cluster centered on the impurity and embedded in a 
field representing the rest of the host lattice. This cluster approach is the
one chosen in the present study, and has been used in the past
to study the geometrical and optical properties
of doped crystals. \cite{Che81,Win87,Kun88,Bar88,Lua89a,Lua89b,Sei91,And91,Isl92,Lua92,Miy93,Sca93,Lua93,Vis93,And93,Flo94,Bel94,Ber95,Pas95,Gry95,Llu96,Sei96,Ber96a,Ber96,Riv98,Agu98,Bar99}
The cluster (active space) can be studied by using standard
quantum-mechanical methods. The rest of the crystal (environment) can be 
described in several ways. In the simplest and most frequently used approach,
the environment is simulated by placing point charges on the
lattice sites, but
this procedure has to be improved in order to obtain a
realistic description of the lattice distortions around the impurity.
\cite{Bar88,Lua89a,Lua89b,Sei91,Lua92,Lua93,Flo94,Pas95,Llu96,Sei96,Agu98}
Model potentials have been developed to represent the effects of the
environment on the active cluster, that include attractive and repulsive
quantum-mechanical
terms aside from the classical Madelung term, \cite{Huz87} but a problem
still remains: the large computational cost of conventional molecular orbital
(MO) calculations prevents from performing an exhaustive geometrical relaxation of the
lattice around the impurity. In the most accurate MO calculations,
\cite{Sei91,Pas95,Llu96,Sei96,Ber96,Bar99} only the positions of the ions in the
first shell around the impurity are
allowed to relax. However,
geometrical relaxations far beyond the first shell of
neighbors can be expected.
In fact, recent semiempirical simulations of solids, 
\cite{Zha93,Zha94,Isl94,Isl95,Say95,Akh95,Exn95,Cat98a,Cat98b} 
performed employing
phenomenological potentials, \cite{Sto81,Har90,Wil93,Wil96,Gal96,Gal97}
have shown the importance of considering
appropriate large-scale lattice relaxations in the study of a variety of
intrinsic and extrinsic defects in ionic crystals. As we will show below, 
the systems
under study in this paper can not be properly described by simply considering
an expansion of the first shell of neighbors around the impurity.

In this contribution we report theoretical calculations of the
lattice distortions induced by Ag$^-$ and Cu$^-$ substitutional impurities in 16
alkali halide crystals with the rock-salt structure, 
namely all those noncontaining
cesium.
For this purpose we use the
{\em ab initio} Perturbed Ion (PI) model,
\cite{Lua90a,Lua90b,Pue92,Lua92a,Lua93a} which circunvents the
problems mentioned above: (a) The active cluster is embedded in an environment
represented by the {\em ab initio} model potentials of Huzinaga 
{\em et al.} \cite{Huz87}; (b) The computational simplicity of the PI model
allows for the geometrical relaxation of several coordination shells around the
impurity. \cite{Lua92,Lua93,Agu98} Moreover, it allows us to study a whole family of
systems in order to look for systematic trends that might be useful in
later theoretical studies of doped crystals similar to those here considered.

The remainder of this paper is organized as follows: 
In Section \ref{aguado:theory}
we describe the active cluster which has been used to model the doped systems.
In Section \ref{aguado:results} we 
present and discuss the results of the calculations, and Section 
\ref{aguado:summary} summarizes
the main conclusions.

\section{Cluster model} 
\label{aguado:theory}

The {\em ab initio} Perturbed Ion model is a particular application of the
theory of electronic separability of Huzinaga and coworkers
\cite{Huz71,Huz73} to ionic solids, 
in which the basic building blocks are reduced to single ions.
The PI model was first developed for perfect crystals. \cite{Lua90a}
Its application to the study of impurity centers
in ionic crystals has been described in refs. \onlinecite{Lua92,Lua93,Agu98}, 
and we refer to those papers for a full account of the method. 
In brief, an active cluster containing the impurity is considered, and
the Hartree-Fock-Roothaan (HFR) equations \cite{Roo63} for 
each ion in the active
cluster are solved in the field of the other ions.
The Fock operator includes, apart from the usual intra-atomic terms, an
accurate quantum-mechanical crystal potential
and a lattice projection operator which accounts for the energy contribution
due to the overlap between the wave functions of the ions. \cite{Fra92}
The atomic-like HFR solutions are used to describe the ions in the 
active cluster
in an iterative stepwise procedure.
The wave functions of the lattice ions external to
the active cluster are taken from a PI calculation for the perfect crystal
and are kept frozen during the embedded-cluster calculation.
Those wave functions are explicitely considered for ions 
up to a distance $d$ from the 
center of the
active cluster such that the quantal contribution from the most distant frozen
shell to the effective cluster energy is less than 10$^{-6}$ hartree.
Ions at distances 
beyond $d$ contribute to the effective energy of the active cluster just
through the long-range Madelung interaction, so they are represented by point
charges.
At the end of the calculation, the ionic wave functions
are selfconsistent within the active cluster and consistent with the frozen
description of the rest of the lattice.  The intraatomic Coulomb correlation,
which is neglected at the Hartree-Fock level, is computed as a correction by
using the Coulomb-Hartree-Fock (CHF) model of Clementi. \cite{Cle65,Cha89}

In a previous work\cite{Agu98} we employed several active clusters of
increasing size and with different embedding schemes to describe the
scintillator system Tl$^+$:NaI. That study was undertaken in order to
find the necessary requirements that a cluster model has to fulfill in order
to describe properly a doped crystal. Here we just describe the best cluster
model between those studied in ref. \onlinecite{Agu98}. This active
cluster, shown in Figure 1, has 179 ions which correspond to the central
impurity (Ag$^-$ or Cu$^-$) plus twelve coordination shells. Those ions are
further split up into two subsets. One is formed by the central impurity plus
the first four coordination shells, having a total of 33 ions, and
both the wave functions and
positions of the ions in this subset are allowed to relax. The lattice positions
of the other 146 ions of the active cluster
are held fixed during the calculations but their
wave functions have been selfconsistently optimized. 
This is done so that the connection between the
region where distortions are relevant and the rest of the crystal
is as smooth as possible. In our previous study \cite{Agu98}
we showed how an unphysically abrupt connection between those two regions 
fails in describing properly the lattice 
distortions induced by the impurity. The ions
in the interface region can respond to those distortions by selfconsistently 
adapting their wave functions to the new potential, and thus contribute to
build a more realistic (selfconsistent) environment.
The geometrical relaxation around the impurity
has been performed by allowing for the independent
breathing displacements of each shell of ions, and minimizing the total
energy with respect to those displacements
until the effective
cluster energies are converged
up to 1 meV.
A downhill simplex algorithm \cite{Wil91} was
used.
For the ions we have used large STO basis sets, all taken from Clementi-Roetti
tables.
\cite{Cle74}

The cluster used in this work has been shown to be self-embedding consistent
for NaI
in our previous work.\cite{Agu98} By this we mean that if the pure crystal is
represented by this cluster model (that is, if the central impurity is replaced
by the halogen ion corresponding to the pure crystal), the results of the
cluster model calculations closely agree with those from a PI calculation for
the pure crystal, where all cations (or anions) are equivalent by translational
symmetry. The same is true for all the family of alkali halide crystals
considered here. Nevertheless, the self-embedding consistency is never complete.
In order to supress systematic errors from the distortions calculated with
the cluster method,
the radial displacements of each shell have been calculated using the
following formula:
\begin{equation}
\Delta R_i = R_i(Imp^-:AX) - R_i(X^-:AX),
\end{equation}
where R$_i$ (i=1, 2, 3, 4) refer to the radii of the first, second, third, and
fourth shells around the impurity in the AX crystal, 
A = Li, Na, K, Rb, X = F, Cl, Br, I, and
Imp$^-$ = Ag$^-$, Cu$^-$. 
Thus both systems (pure and doped crystals) are treated in eq. (1) on equal
foot with the cluster model,
and not with different methodologies, and the calculated distortions are
free from that potential source of error. Also, in order to have the correct
Madelung potential at the impurity site, the calculations have been performed
by employing the experimental lattice constants\cite{Ash76} to describe the
geometrically frozen part of the crystals.

The only terms omitted in our
description are the dispersion terms (coming from interatomic correlation) and
relativistic effects for the heavy ions. Although the importance of both effects
increase with atomic number, they are not crucial for the structural properties
of the systems studied here. Specifically, Mart\'\i n Pend\'as {\em et al.}
\cite{Mar97} have shown that the PI method gives lattice constants and
bulk moduli in close agreement with experimental results for all alkali halides.
The properties of these crystals under the influence of an applied external 
pressure, a situation where the importance of
interatomic correlation effects increases, are also properly reproduced.

\section{Results and discussion}
\label{aguado:results}

The calculated distortions, collected in Table I, are the
main quantitative result from our study. For visualization
of the trends, however, it is better to display the results in
a figure, and this is done in Fig. 2, where we have plotted the distortion of
each of the four shells in terms of the empirical
cationic radii extracted from ref.
\onlinecite{Ash76}. Those points corresponding to the same anion have been
joined with a line to guide the eye. The figure 
contains only the results for
Ag$^-$ because the trends are the same in the case of the Cu$^-$
impurity. Next we describe, shell by shell, the general trends in Fig. 2:

{\em First shell.} This shell is formed by 6 cations in ($\frac{1}{2}$,0,0)
crystallographic sites, and undergoes an
expansion, as might be expected from the larger size of Ag$^-$
compared to the halogen anions. The impurity anion pushes the neighbor cations
to make room for itself in the lattice. The expansion is substantial, with
percentage values between 9 and 15 \%.
In the F salts that expansion is larger the
larger the cation size, 
but this trend is violated in the Cl, Br, and I crystals.
If we fix the cation, for K and Rb salts the expansion is larger
the smaller the anion (notice that the anion size increases in the order F$^-$,
Cl$^-$, Br$^-$, I$^-$). This rule is inverted in the case of
Li salts, whereas Na salts constitute an intermediate
case. It should be recognised, nevertheless, that the expansion is almost
independent of the halogen element in the Li and Na salts.

{\em Second shell.} The displacement of the second shell, formed by 12 anions at
($\frac{1}{2}$,$\frac{1}{2}$,0) positions, is
always a small contraction. If we fix the anion, 
the contraction is larger (absolute
value) the larger is cation size. If the cation is fixed, for Na, K and Rb salts
the contraction is larger the smaller the anion size. Again this trend is
inverted for Li salts.

{\em Third shell.} This shell, formed by 8 cations at
($\frac{1}{2}$,$\frac{1}{2}$,$\frac{1}{2}$) sites, experiences a small
expansion. If we fix the anion, the expansion is smaller the larger the
cation size. If the cation is fixed, there is not a definite trend.
In the case of Li and Na salts the expansion increases with the
anion size (LiI is an exception). This trend is partially inverted in the case
of Rb salts, and for K salts all the expansions are almost
identical.

{\em Fourth sell.} This shell, formed by 6 anions at (1,0,0) 
positions, experiences an expansion. That expansion increases with cation size
if we fix the anion. If the cation is fixed, in K and Rb salts the expansion
is smaller the larger the anion size. In Na salts, NaI is again an
exception to this general rule, whereas in Li salts the expansion is
almost constant.

In the following we try to find some rationalization for the calculated trends.
The working rule still
in use nowadays stating that the expansion 
of the first coordination shell can be
approximated by the difference between the ionic radii of the impurity and the
substituted ion is somewhat misleading. First of all, the ions in a
crystal are not hard spheres, but weakly overlapping soft spheres,
so one cannot
use values for the ionic radii of ions in vacuum in order to {\em predict} 
lattice distortions accurately. 
In particular, the size of an anion may vary in a
nonnegligible way from crystal to crystal.
To investigate this, we show
$(<r^2>)^{1/2}$ for F$^-$ in fluoride crystals in Table II, 
where the expectation value is taken 
over the outermost orbital of
the anion (2p), and is 
calculated from the crystal-consistent ionic wave
functions obtained through a PI calculation on the pure crystals. 
r$_X$ = $(<r^2>)^{1/2}$ can be taken as a rough measure of the anion size.
We also show the analogous quantity
for the 5s orbital of the Ag$^-$ impurity in
fluorides.
The size of the F$^-$ anion varies by a maximum of 2 \%, small
compared with
the size variation of the Ag$^-$ anion (7 \%). Ag$^-$ is more compressible than
the halogen anions because its outer electronic shell is an s-shell, while it is
a p-shell for the halogens. The same can be said of all the
halogens. As the size of the cation decreases, Ag$^-$ is more
compressed by the crystal environment. This shows that the standard ionic radii
cannot be used for predicting distortions, 
because their values are a genuine {\em output}
of the selfconsistent process. Nevertheless, though they are not useful for
accurate predictions, physical insight tells us that the distortions should be
correlated with the size of the ions, in the sense that one always expect that
larger impurities induce larger distortions. These should be useful at 
least at a qualitative level.

In Table III we show the differences 
\begin{equation}
\delta =r(Ag^-:AX)-r_X(X^-:AX),
\end{equation}
where r(Ag$^-$:AX) is the radius of the Ag$^-$ impurity in the AX crystal, and
r$_X$(X$^-$:AX) is the radius of the halogen anion. The differences
are expected to be related with the first-shell expansions
$\Delta R_1$. The values of $\delta$ do not show quantitative agreement with
those of $\Delta R_1$. If the anion is fixed, $\delta$ increases with
the cation size, which is consistent with the main trend in $\Delta R_1$ of
fluorides, but 
does not explain the behavior of other halides. The main
trends discussed when the cation is fixed are reproduced for the K and Rb
crystals but not for the others. From Figure 1 we can see that the exceptional
crystals, concerning the trends in $\Delta R_1$, are LiCl, LiBr, LiI, and NaI.
The peculiar feature of
those four cases is that anions are
much larger than cations: the ratios r$_A$/r$_X$ are the smallest in the family
of alkali halides. In a study of the structures of small alkali halide
clusters, \cite{Agu97} it was found that materials with small r$_A$/r$_X$ 
have a cluster growing
pattern different from the rest. Specifically, those systems showed a marked
tendency towards ring-like structures, whereas the others adopt fragments
of the rocksalt lattice as their minimum energy structures. The
main reason is that anion-anion repulsions are much more important when the
ratio r$_A$/r$_X$ is small. In a recent study, A. Mart\'\i n Pend\'as
{\em et al.} \cite{Mar98} have applied the atoms--in--molecules (AIM) theory of
Bader\cite{Bad90} to study the topology of the electron density in crystals.
They found that the whole group of alkali halides with rocksalt structure can
be divided up into three topological families, 
called R$_1$, B$_1$, and
B$_2$ in that paper, and that the ionic radii are the topological organizers. 
The R$_1$ family
contains KCl, NaF, KF, RbF, RbBr, RbCl and all the cesium halides. 
The B$_1$ family
contains KI, KBr, LiF, RbI, NaBr, NaCl and NaI, and the B$_2$ family contains
LiCl, LiBr and LiI. There are constant r$_A$/r$_X$ lines that isolate each
family. The largest values of r$_A$/r$_X$ are found for family R$_1$ and the
smallest for family B$_2$, with intermediate values for B$_1$.
NaI is so near the B$_2$ region that it is not surprising that it behaves in
Fig. 2 like the elements of the B$_2$ family. In the AIM theory
the critical points of the electron density scalar field
are classified as nuclei, bond points, ring points and cage
points.\cite{Bad90} When a bond point is found between two nuclei, a bond is
established between the corresponding atoms.
In the R$_1$
family there are just anion-cation bonds.\cite{Mar98} 
We have found that for the (undoped) crystals
of this family, the anion-anion overlap
is at least one
order of magnitude smaller than the anion-cation overlap. In the crystals of the
B$_1$ and B$_2$ families 
there are bond critical points between anions, so that the effective local
coordination is 6 for cations and 18 for anions (6 anion-cation and 12 
anion-anion bonds).\cite{Mar98} In the B$_1$ crystals (excepting NaI)
we have found
that the anion-anion overlap is smaller but of the same order of magnitude than
cation-anion overlap. In the B$_2$ crystals, and also in NaI,
anion-anion overlap is the largest contribution to the
repulsive interactions, and thus anion-cation contacts are
less important.
The cations of the B$_2$ family occupy the interstitial holes left 
in the anionic fcc
sublattice. It is then not surprising that when the cation-anion overlap is not
so important, the expansion of the cation shell is an exception to the general
trends; it is, in fact, nearly constant for the B$_2$ family.

Let us turn to discuss the distortion of the second and third shells.
In all cases, the second shell suffers a contraction of a small magnitude
compared to the large expansion of the first shell. In Figure 1 one can see
that the radial outward motion of the first cation shell is not going to affect
much the positions of the twelve anions of shell 2, so it becomes understandable
that the anions of that shell move little. The small contraction of the second shell
optimizes the Madelung energy around the impurity and also serves to pack more
efficiently the ions in response to the outward motion of the cations.
The quantitative trend of
that contraction is understood with reference to the three topological families
discussed in the previous paragraph and their relation to the anion-anion
overlap: the contraction is largest for those crystals where the anion-anion
overlap is small (R$_1$ crystals), intermediate when that
overlap begins to count (B$_1$ crystals), and finally, it is lowest
for the B$_2$ crystals, where anion-anion contacts are important. This
explains the trends observed: if the anion is fixed, the 
contraction is larger the larger the cation size, because the Ag$^-$-X$^-$
overlap decreases with increasing cation size. On the other hand,
if the cation is fixed,
in Rb, K, and Na salts the contraction decreases with increasing anion size, 
because the Ag$^-$-X$^-$ overlap increases with anion size. But in 
Li salts the trend is
inverted because Ag$^-$-X$^-$ overlaps decrease with anion
size. The distortion of the
third shell is not directly related to the introduction of the impurity, 
as the overlap between the cations of that shell and Ag$^-$ is very small (see Fig. 1). 
The expansion
of this shell appears to be again of a purely electrostatic origin.
The relative values
of the displacements (R$_3$ - R$_3^{crystal}$)/R$_3^{crystal}$ are very small,
less than 1 \% except in LiF. 

The $\Delta R_4$ displacements, always an expansion, proceed along the
same crystallographic direction as the $\Delta R_1$ displacements. 
Thus, the
expansion is clearly induced by
the expansion of the first shell. The relative displacements 
(R$_4$-R$_4^{crystal}$)/R$_4^{crystal}$ adopt values between 1 \% and 5\%,
compared to values of 9--15 \% for (R$_1$-R$_1^{crystal}$)/R$_1^{crystal}$.
These numbers indicate that the expansion of the (ninth) shell formed by six cations at 
($\frac{3}{2}$,0,0) is not expected to be higher than 1 \%.
In the K and Rb salts $\Delta R_4$ is larger
when $\Delta R_1$ is larger, while in Li and Na salts $\Delta R_4$ and
$\Delta R_1$ are both almost independent of the anion, so the displacements of 
the first and fourth
shells are correlated.
If the anion is fixed,
the expansion increases with the cation size, and no
special behaviour is observed in the cases of LiCl, LiBr, LiI and NaI. 

We conclude that the lattice relaxation around the substitutional impurity in
the alkali halides involves the concerted movement of
several coordination shells. However, it is not yet clear from the results
presented up to this point whether the lattice relaxations of the second, third and
fourth shells have a substantial influence on the energy of formation
of the defect. At low pressure and temperature conditions, the formation of the
impurity centers should be discussed in terms of the internal energy difference
for the exchange reaction\cite{Flo94}
\begin{equation}
(X^-:AX)_s + Imp^-_g \rightleftharpoons (Imp^-:AX)_s + X^-_g,
\end{equation}
where the $s$ and $g$ subindexes refer to solid and gas phases, respectively.
In order to establish the importance of the relaxation of the lattice
beyond the first coordination shell, we have calculated the formation
energy $\Delta$H of the defects by employing two different models for the
active cluster, shown in Fig. 
1: one of them is that formed by 179 ions described in section II;
the other includes just four coordination shells around the central ion
(a total of 33 ions), and only the
positions of the six ions in the first coordination shell are allowed to relax.
The results are shown in Table IV. We see that according to the small cluster
model the energy of formation of the defects is always positive, that is none 
of the impurity centers are stable centers. Enlarging the cluster
size to include a selfconsistent treatment of 179 ions and extending the 
lattice relaxation up to the fourth coordination shell induces a huge 
stabilization of all the impurity centers.
The trend of $\Delta$H is simple. For a given impurity (Ag$^-$ or Cu$^-$) the
heat of formation decreases by increasing the atomic number of the cation
(alkali) or of the anion (halogen). The positive value of $\Delta$H in the
calculation for the small cluster is understandable: the impurity simply pushes
its neighbor cations producing a high elastic strain energy since the rest of
the lattice is not allowed to respond. In the second model three more shells
are allowed to move in response to that initial stress and the relaxation lowers
the elastic energy so much that the electronic contributions turn $\Delta$H
negative in all cases except Ag$^-$:LiF and
Ag$^-$:NaF ($\Delta$H is nearly zero in Ag$^-$:LiCl). It is useful to notice that $\Delta$R$_1$,
the displacement of the first shell, is lower for the first cluster model
compared to the second. This means that due to the constraints imposed by the
first model the atoms of the first shell are unable to reach their preferred
equilibrium positions in the presence of the impurity, a fact that is
consistent with the large calculated $\Delta$H. The change of sign in
$\Delta$H can be interpreted as suggesting that most of the elastic relaxation
of the lattice has been accounted for and that allowing for the elastic
relaxation of more shells will have a minor effect. The two cases with a
positive $\Delta$H, Ag$^-$:LiF and Ag$^-$:NaF, are still intriguing. These two crystals have
the smallest lattice parameters within the whole family studied here, and it is
conceivable that the elastic effects will be largest. The question if the
relaxation of more coordination shells is able to stabilize those two systems
deserves further investigation.

As indicated above the difference in the heats of formation given by the two
models, $\Delta$H(model 2) - $\Delta$H(model 1), gives a measure of the lowering
of elastic strain when more coordination shells are allowed to relax. From 
Table IV one can verify that, for a given crystal, this energy is essentially
independent of the impurity, while both $\Delta$H(model 1) and 
$\Delta$H(model 2) depend on the impurity. This confirms our interpretation
of the effect of allowing for the elastic relaxation of several shells: that
relaxation is mainly a host effect.

\section{Summary}
\label{aguado:summary}

We have reported a study of the local lattice distortions
induced by substitutional Ag$^-$ and Cu$^-$ impurities in 
the family of alkali halide crystals excepting those containing cesium.
For this purpose, the {\em ab initio} Perturbed Ion (PI) model has been used.
A large active cluster of 179 ions, 
embedded in an accurate quantum environment representing
the rest of the crystal, has been studied.
The local distortions
obtained extend beyond the first shell of neighbors in all cases.
Thus, the assumptions frequently employed in impurity calculations,
which consider the active space as formed by the central impurity plus its
first coordination shell only, should be taken with some care.
Distortion trends have been identified and discussed. The first coordination 
shell (cations) around the impurity 
experiences an expansion as a consequence of
the larger size of the impurity anion compared to the halogens. That expansion
is larger for the Ag$^-$ than for the Cu$^-$ impurity, also because the first
anion is larger than the second. The trends can be qualitatively explained by
considering the difference in size between the impurity and the
substituted anion in all cases except in those crystals with a very small size
ratio between cation and anion. In those cases, that is for LiCl, LiBr, LiI
and NaI, anion-anion
contacts are important.
Those four materials have been found to exhibit special 
behavior in a number of previous studies involving crystals and clusters.
The contraction of the second shell as well as the expansion of the
third shell are small and arise from a combination of electrostatic and packing
origins.
The fourth shell experiences a substantial
expansion as a consequence of the direct pushing induced by the
expansion of the first shell. 
The analysis of the energies of formation of the defects clearly shows that
elastic relaxation of several coordination shells around the impurity is
necessary in the modeling of these materials since this affects even the sign
of the energy of formation.

$\;$

$\;$

{\bf Acknowledgements:} Work supported by DGES (PB98-0345) and Junta de
Castilla y Le\'on (VA28/99). 
A. Aguado is grateful to University of Valladolid for financial support.
We thank the suggestions of one referee.

{\bf Captions of Tables}

$\;$

{\bf Table I.}
Radial displacements $\Delta R_i$ (see eq. (1)), in \AA, of the 
first four shells of ions around the
silver and copper impurities. 

{\bf Table II.} 
$(<r^2>)^{1/2}$, where the expectation values are taken over the outermost 
orbital of the F$^-$ anion
in pure alkali fluorides and of the silver anion in Ag$^-$-doped alkali
fluorides. All quantities in \AA.

{\bf Table III.}
Difference of radii between Ag$^-$ and the substituted anion (see eq. (2)),
in \AA.

{\bf Table IV.}
Formation energies (in eV) of copper and silver substitutional centers in different alkali halide
host lattices, calculated employing two different models for the active cluster.
First row:
the active cluster contains 33 ions, and only the positions of the ions in the
first coordination shell are allowed to relax. Second row: the active cluster
contains 179 ions, and the positions of the ions in the first four coordination
shells are allowed to relax.

$\;$

{\bf Captions of Figures}

$\;$

{\bf Figure 1.}
The active cluster (ImpA$_{92}$X$_{86}$)$^{5+}$
employed to represent the region around the impurity, where Imp=Ag, Cu; 
light spheres
are cations and dark spheres anions. The core of the cluster, formed by the
four first coordination shells, which are allowed to breath, is also 
indicated separately.

{\bf Figure 2.}
Shell distortions $\Delta$R$_i$ (i=1,2,3,4) around Ag$^-$, 
plotted as a function of cation
size.

\newpage

\onecolumn[\hsize\textwidth\columnwidth\hsize\csname
@onecolumnfalse\endcsname

\begin{table}[t]
\begin{center}
\begin{tabular} {|c|c|cc|c|cc|c|cc|c|cc|}
\hline
& Crystal & Ag$^-$ & Cu$^-$ & Crystal & Ag$^-$ & Cu$^-$ & Crystal & Ag$^-$ & Cu$^-$ & Crystal & Ag$^-$ & Cu$^-$ \\
\hline
& LiF & 0.288 & 0.267 & LiCl & 0.294 & 0.290 & LiBr & 0.299 & 0.294 & LiI & 0.308 & 0.302\\
&    & -0.0416 & -0.0371 &  & -0.0439 & -0.0403 &  & -0.0478 & -0.0467 & & -0.0492 & -0.0484 \\
&     & 0.0418 & 0.0397 &    & 0.0439 & 0.0425 &    & 0.0462 & 0.0452 &   & 0.0457 & 0.0446 \\
&     & 0.0687 & 0.0619 &    & 0.0622 & 0.0617 &    & 0.0642 & 0.0636 &   & 0.0732 & 0.0732 \\
\hline
& NaF & 0.308 & 0.292 & NaCl & 0.289 & 0.285 & NaBr & 0.290 & 0.284 & NaI & 0.299 & 0.288 \\
&     & -0.0660 & -0.0621 &  & -0.0628 & -0.0620 &  & -0.0600 & -0.0591 & & -0.0552 & -0.0540 \\
&     & 0.0336 & 0.0336 &    & 0.0350 & 0.0350 &    & 0.0388 & 0.0383 &   & 0.0401 & 0.0404 \\
&     & 0.121 & 0.114 &      & 0.107 & 0.100 &      & 0.0867 & 0.0843 &   & 0.0915 & 0.0797 \\
\hline
& KF  & 0.364 & 0.355 & KCl  & 0.326 & 0.315 & KBr  & 0.311 & 0.300 & KI  & 0.284 & 0.269 \\
&     & -0.0778 & -0.0695 &  & -0.0739 & -0.0686 &  & -0.0719 & -0.0709 & & -0.0709 & -0.0699 \\
&     & 0.0319 & 0.0314 &    & 0.0316 & 0.0316 &    & 0.0313 & 0.0307 &   & 0.0318 & 0.0312 \\
&     & 0.224 & 0.220 &      & 0.205 & 0.200 &      & 0.172 & 0.166 &     & 0.145 & 0.136 \\
\hline
& RbF & 0.416 & 0.408 & RbCl & 0.351 & 0.343 & RbBr & 0.332 & 0.318 & RbI & 0.300 & 0.281 \\
&     & -0.0828 & -0.0736 &  & -0.0749 & -0.0731 &  & -0.0728 & -0.0728 & & -0.0727 & -0.0716 \\
&     & 0.0307 & 0.0292 &    & 0.0268 & 0.0262 &    & 0.0261 & 0.0261 &   & 0.0286 & 0.0286 \\
&     & 0.254 & 0.249 &      & 0.226 & 0.220 &      & 0.195 & 0.187 &     & 0.166 & 0.156 \\
\hline
\end{tabular}
\end{center}
\end{table}

\begin{table}[t]
\begin{center}
\begin{tabular} {|c|c|c|c|}
\hline
& Crystal & $(<r^2>)^{1/2}$(F$^-$) & $(<r^2>)^{1/2}$(Ag$^-$) \\
\hline
& LiF & 0.698 & 1.303 \\
& NaF & 0.704 & 1.343 \\
& KF & 0.710 & 1.380 \\
& RbF & 0.712 & 1.401 \\
\end{tabular}
\end{center}
\end{table}

\begin{table}[t]
\begin{center}
\begin{tabular} {|c|c|c|c|c|c|c|c|c|}
\hline
& Crystal & $\delta$ & Crystal & $\delta$ & Crystal & $\delta$ & Crystal & $\delta$ \\
\hline
& LiF & 0.605 & LiCl & 0.343 & LiBr & 0.273 & LiI & 0.162 \\
& NaF & 0.639 & NaCl & 0.388 & NaBr & 0.328 & NaI & 0.209 \\
& KF & 0.670 & KCl & 0.401 & KBr & 0.368 & KI & 0.241 \\
& RbF & 0.689 & RbCl & 0.460 & RbBr & 0.378 & RbI & 0.251 \\
\end{tabular}
\end{center}
\end{table}

\begin{table}
\begin{center}
\begin{tabular} {|c|c|cc|c|cc|c|cc|c|cc|}
\hline
& Crystal & Ag$^-$ & Cu$^-$ & Crystal & Ag$^-$ & Cu$^-$ & Crystal & Ag$^-$ & Cu$^-$ & Crystal & Ag$^-$ & Cu$^-$ \\
\hline
& LiF & 16.79 & 15.41 & LiCl & 11.62 & 11.04 & LiBr & 11.63 & 11.02 & LiI & 9.76 & 9.30\\
&    & 1.29 & -0.22 &  & 0.02 & -0.57 &  & -0.21 & -0.64 & & -0.61 & -1.10 \\
\hline
& NaF & 13.38 & 12.44 & NaCl & 11.20 & 10.63 & NaBr & 10.38 & 9.91 & NaI & 8.68 & 8.52 \\
&     & 0.45 & -0.55 &  & -0.02 & -0.63 &  & -0.24 & -0.87 & & -0.85 & -1.01 \\
\hline
& KF  & 11.25 & 10.42 & KCl  & 9.85 & 9.32 & KBr  & 9.08 & 8.69 & KI  & 7.82 & 7.55 \\
&     & -0.33 & -1.14 &  & -0.84 & -1.28 &  & -1.04 & -1.26 & & -1.11 & -1.20 \\
\hline
& RbF & 10.37 & 9.60 & RbCl & 9.19 & 8.73 & RbBr & 8.76 & 8.39 & RbI & 7.56 & 7.31 \\
&     & -0.57 & -1.31 &  & -0.91 & -1.38 &  & -0.97 & -1.35 & & -1.09 & -1.35 \\
\hline
\end{tabular}
\end{center}
\end{table}

\newpage

\begin{figure}
\psfig{figure=nacl33.epsi}
\end{figure}

\newpage

\begin{figure}
\psfig{figure=nacl179.epsi}
\end{figure}

\newpage

\begin{figure}
\psfig{figure=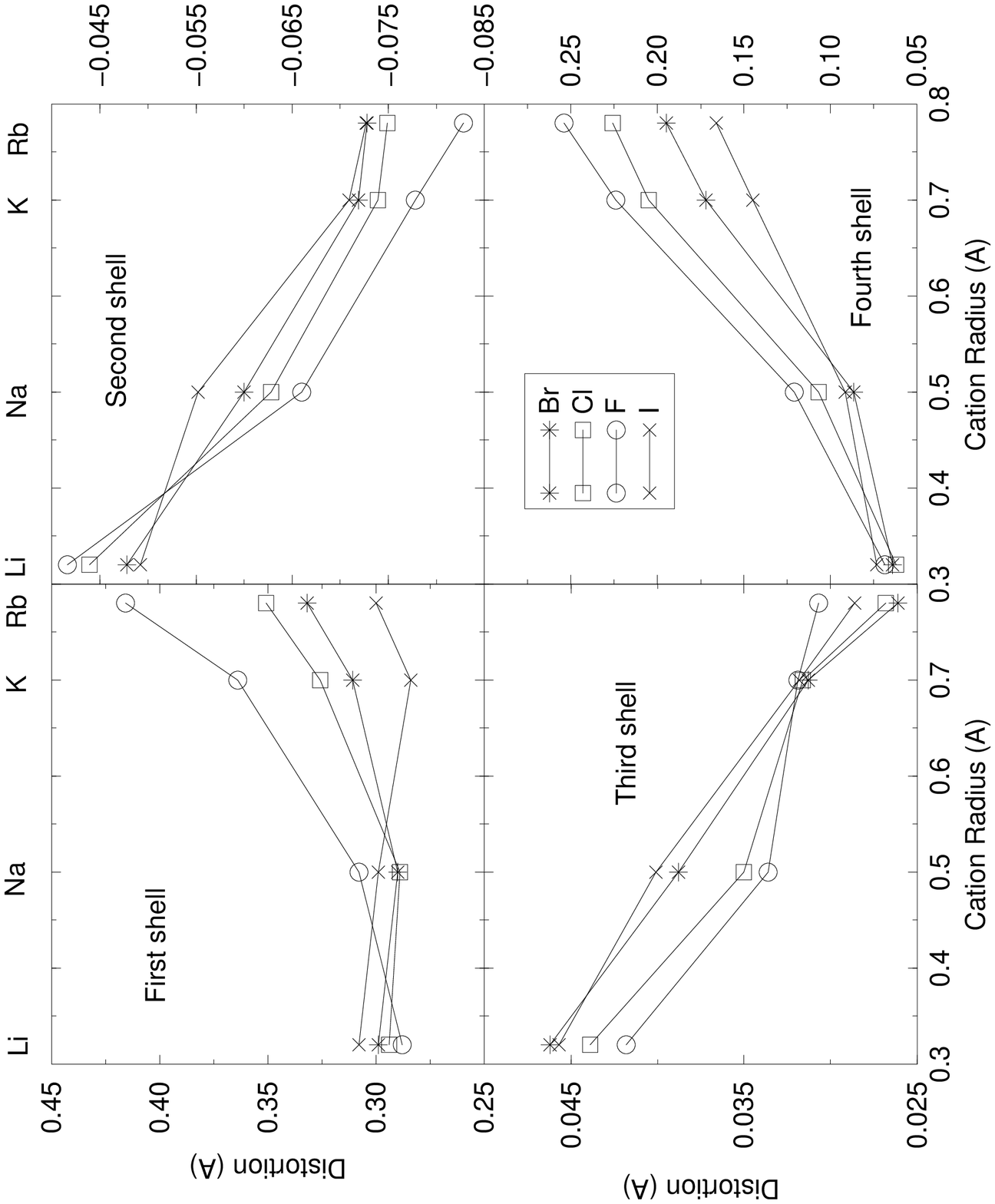}
\end{figure}

\end{document}